\documentstyle[epsfig]{aipproc}
\newcommand\euro{{\sffamily C%
\makebox[0pt][l]{\kern-.70em\mbox{--}}%
\makebox[0pt][l]{\kern-.68em\raisebox{.25ex}{--}}}}

\begin{document}
\pagestyle{plain}

%\begin{flushright}

%{\small
%CLIC Note 418\\
%LCC-0009\\
%SLAC-PUB-8725
%}

%\end{flushright}

\title{Test Beams and Polarized Fixed Target Beams at the NLC}

\author{Lewis Keller, Rainer Pitthan, Sayed Rokni, and Kathleen Thompson 
\footnote[1]{Talk given (R. P.) at the Fifth International Linear Collider Workshop (LCWS2000), 
Fermi National Accelerator Laboratory, Batavia,  Illinois, US, October 24-28, 2000.\\
Work supported by DOE, contract DE-AC03-76SF00515.} 
}
\vskip-20pt
\address{Stanford Linear Accelerator Center\\
Stanford University, California 94309
}
\vskip-20pt
\author{Yury Kolomensky 
}
\vskip-20pt
\address{California Institute of Technology\\
Pasadena, California 91125
}
\vskip-20pt

\maketitle
%\vskip-20pt
\begin{abstract}
A conceptual program to use NLC beams for test beams and fixed target physics is described.
Primary undisrupted polarized beams would be the most simple to use,
but for NLC, the disrupted beams are of good enough quality that they could also be used, after
collimation of the low energy tails, for test beams and fixed target physics.
Pertinent issues are: what is the compelling physics, what are the requirements on beams and running time, 
and what is the impact on colliding beam physics running.
A list of physics topics is given; one topic (M\o ller Scattering) is treated in more depth. 
\end{abstract}
%\vskip-20pt
\section{Introduction}
 
During the last 10 years, colliding beam physics with the SLC/SLD was SLAC's main program.
During most of the SLC colliding beam running, SLAC was also engaged in a highly successful
fixed target program.
The main emphasis was on the spin content of the nucleon (E-142 \cite{slace142} and E-143 \cite{slace143} 
at 30 GeV, and E-154 \cite{slace154} and E-155 \cite{slace155} at 50 GeV), but other fundamental experiments like E-144 also ran \cite{slace144}.
Typically two months per year were set aside from the main program.
Test beams\cite{cavalli93}) were used parasitically for smaller experiments
(E-146, \cite{LPM}) and for detector development, in particular GLAST \cite{glast96}.

This report discusses the possibilities for using the
NLC \cite{nlczdr} primary electron beam for fixed target physics or to produce secondary
hadron test beams.
While dedicated operation of a fixed target program is obviously the most efficient, we will
assume that test beam or fixed target running is done in parallel with colliding beam physics.
Other design conditions are that there must be permitted access to the fixed target
and test beam experimental building during colliding beam operation, there should be useful hadron and
electron yields up to about 80\% of the primary beam energy, and a single beam transport line and
experimental area can be used for both tests and fixed target running.

\begin{figure}[bhtp] % fig 1
\centerline{\epsfig{file=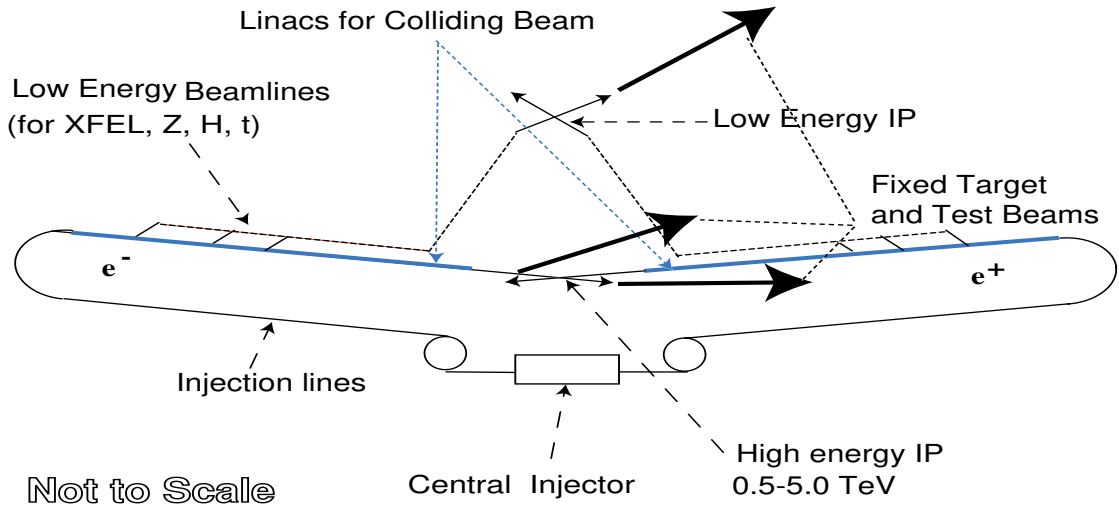,height=3.2in,width=6in}}
\vspace{0pt}
\caption{Schematic layout for NLC. 
Note the linac break-out sections at the 50 GeV point (for an XFEL at 0.1\AA~and/or measurements of the Z), and other energies to measure the Higgs and top.
Test beam or fixed target facilities could be attached behind either of the IR's 
or at the end of the electron linac.
}
\label{fig:GenLayOut}
\end{figure}

Figure \ref{fig:GenLayOut} shows possible locations of such test beam and fixed target beam facilities: they could be at the end of the electron linac (dedicated running only),
or behind either IP following the NLC extraction line.
The scheme shown is based on the current NLC design of a low energy IR (up to 500 GeV center of mass (cms), i.e., 250 GeV per beam), and a high energy IR (up to 1 TeV cms) expandable to 
higher energies.
To keep the option for multi-TeV beams open, the beam path to the high energy IR has minimal bends;  this requires that the linacs have a small angle of $\approx 20$~mrad between them to enable the outgoing bunches to be separated from the incoming beam.

\begin{figure}[t] % fig 2
\centerline{\epsfig{file=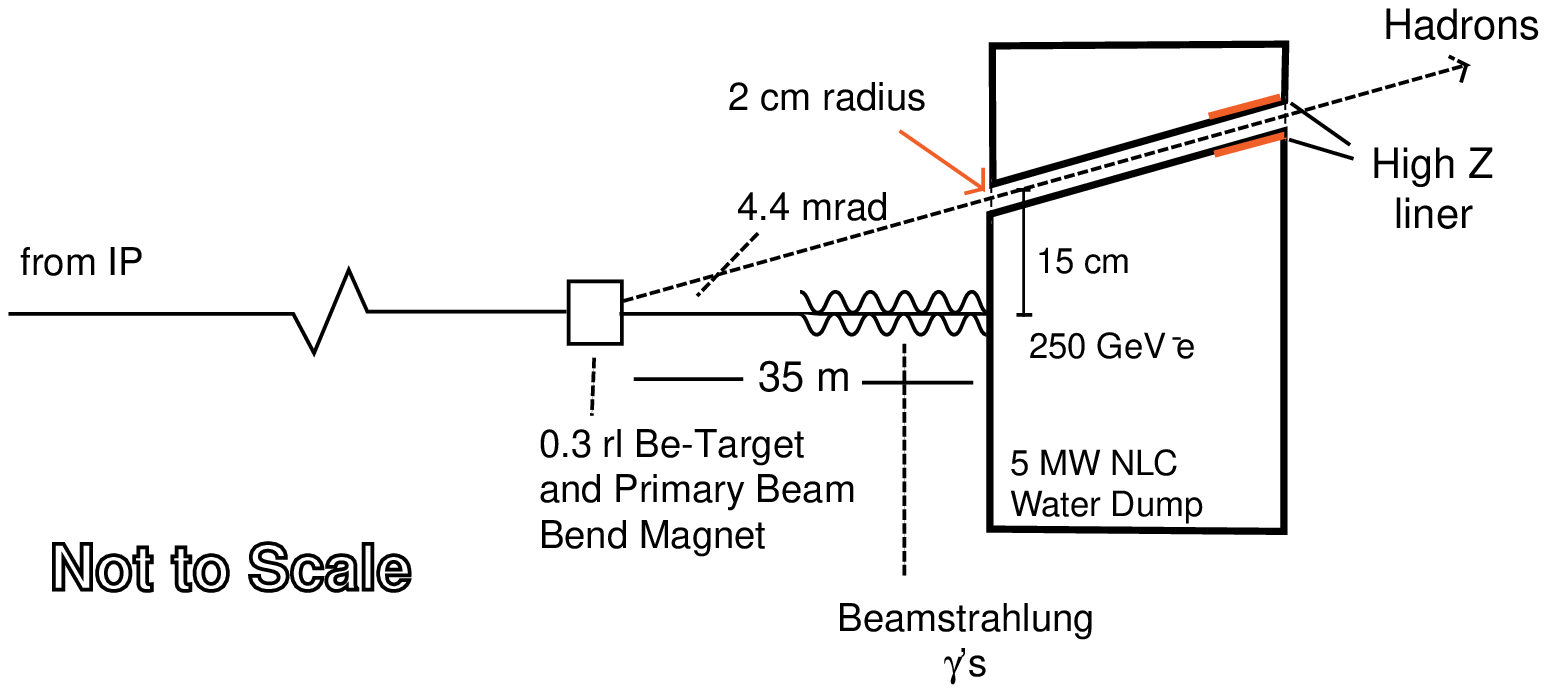,height=3.2in,width=6in}}
\vspace{0pt}
\caption{Schematic layout of the NLC low energy IR extraction line, including the 5~MW water beam dump and the proposed test beam and fixed target take-off.
}
\label{fig:TestBeam}
\end{figure}

Figure \ref{fig:TestBeam} shows
a schematic layout of the NLC low energy IR extraction line, including the 5~MW water beam dump,
and the proposed test beam and fixed target take-off.
For colliding beam physics the beam dump intercepts both the disrupted, primary electrons and the beamstrahlung $\gamma$'s.
For test beams, a 0.3 radiation length Be target is placed in the disrupted beam for production of secondary e$^-$, $\pi^-$, K$^-$, \={p} (or their anti-particles).
To extract the primary beam for fixed target physics, a dipole is placed at the location of the
Be target to bend the beam into the secondary transport line.
The dipole aperture must be large enough so as not to intercept the beamstrahlung photons.
The 5~MW dump (a larger dump will be required for the high energy IR)
does double duty as a collimator of the disrupted beam energy tail, see Section II-B.
Some design requirements of the transport beam line to the test beam/fixed target experimental area are: 
1) able to transport the full primary beam energy, 
2) able to collimate the disrupted beam energy tail within 1\% of the undisrupted energy, 
3) no energy dispersion in the experimental area, 
4) magnet apertures sized to allow useful secondary yields, 
5) zero net bend so that the longitudinal polarization is preserved no matter what beam energy is being used in NLC, 
6) the experimental area has a sufficient transverse displacement from
the straight-ahead NLC dump line so that the muon flux from the 5~MW dump is not a problem.

%%%%%%%%%%%%%%%%%%%555
\section{Test Particle Production and Tail Collimation}
\subsection{Test Particle Production}

\begin{figure}[bhtp] % fig 3
\centerline{\epsfig{file=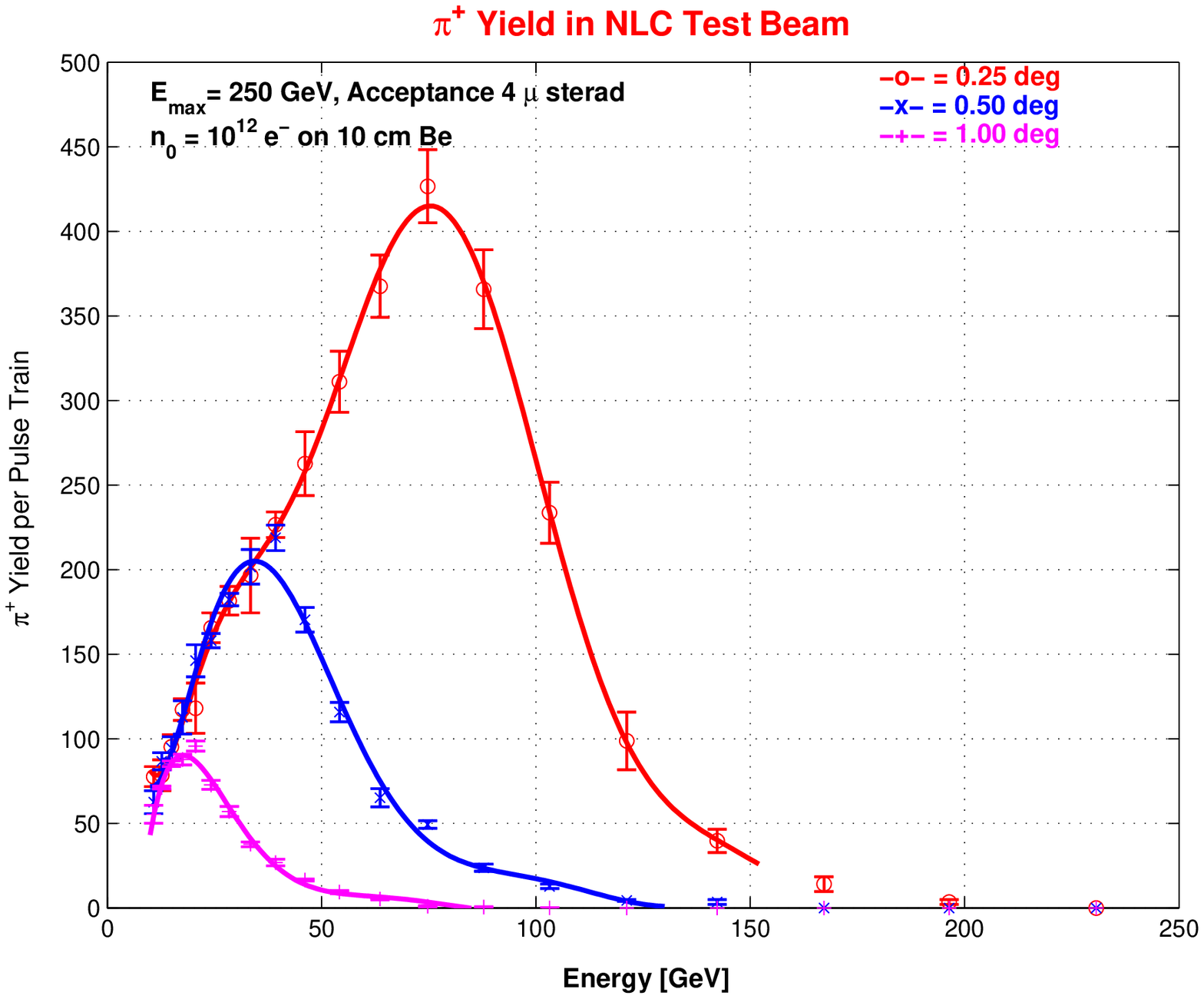,height=2.4in,width=3.2in}
            \epsfig{file=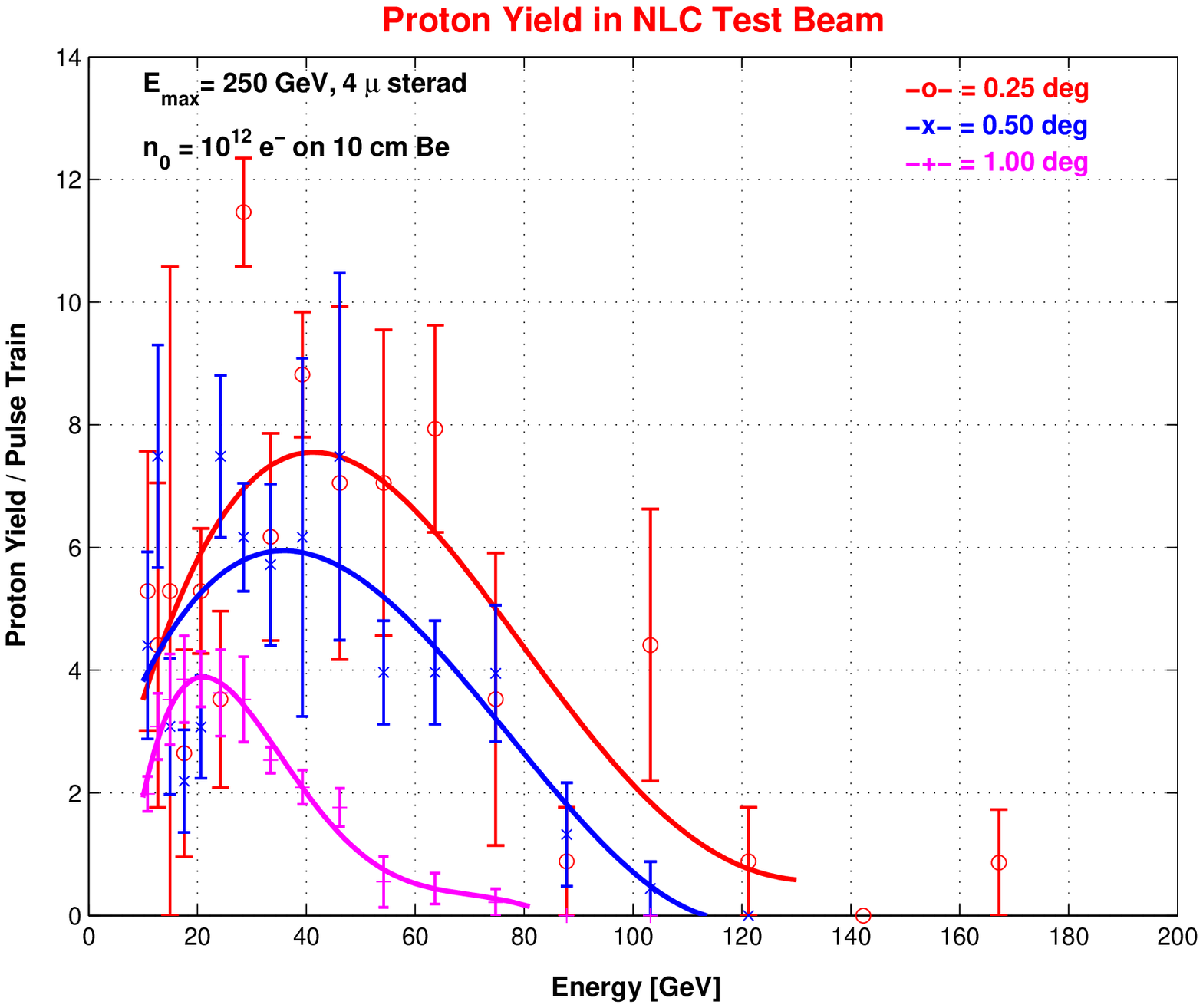,height=2.4in,width=3.2in}
            }
\vspace{0pt}
\caption{Pion and proton yields into 4~$\mu$sr solid angle and $\Delta P/P$ = 4\% per nominal NLC pulse train,
for three angles, calculated with FLUKA [10].
The yield at low energies is sufficient at all angles shown; simulations with more test particles
will be performed in the future to explore the high energy region at larger angles with greater precision.
}
\label{fig:Yield}
\end{figure}

Figure \ref{fig:Yield} shows 
simulations \cite{fluka} of pion and proton yields
 at production angles of 0.25$^\circ$,
0.5$^\circ$ and 1.0$^\circ$ for a 250 GeV electron beam of intensity 10$^{12}$/pulse train on a 0.3 rl
Be target in a secondary beam line of acceptance 4~$\mu$sr and $\Delta P/P$ = 4\%.
The curves show that the yield at 1$^\circ$ is large enough for lower energies,
but that the maximum energy of particles is a function of angle, so smaller angles give more 
latitude in choosing the right yield at higher secondary energies.
Kaon yields are somewhere in between.
Not enough test particles were used in these simulations to determine if enough yield 
is produced at an angle of 1$^\circ$; this will be investigated at a later time.

%%%%%%%%%%%%%%%%%%%%%%%
\subsection{Fixed Target Experiments}

\begin{figure}[bhtp] % fig 4
\centerline{\epsfig{file=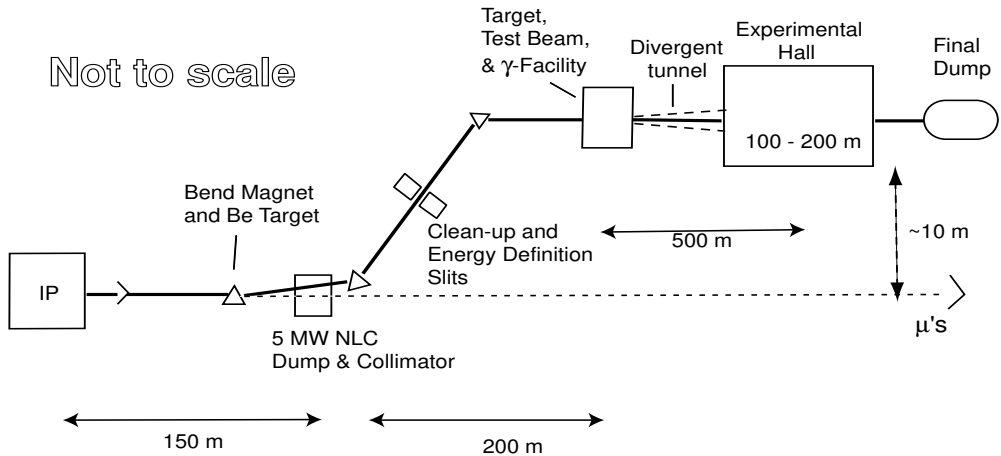,height=2.4in,width=5.2in}}
\vspace{10pt}
\caption{Conceptual layout of the fixed target area.
The target room doubles as a measurement station for the test beam.  
The large energies require small scattering angles and, therefore, large drift sections.
The target and spectrometer locations are separated by a divergent tunnel 
to minimize the excavation costs.
}
\label{fig:FTArea}
\end{figure}

Figure \ref{fig:FTArea} shows the concept of a combined test beam and fixed target facility. 
The beams after the dump are bent and sent through a system of slits. 
These slits will be designed to define the momenta of the 
particles for the test beam application and to clean up the 
remainder of the tails of the disrupted beams for fixed target experiments.

\begin{figure}[bhtp] % fig 5
\centerline{\epsfig{file=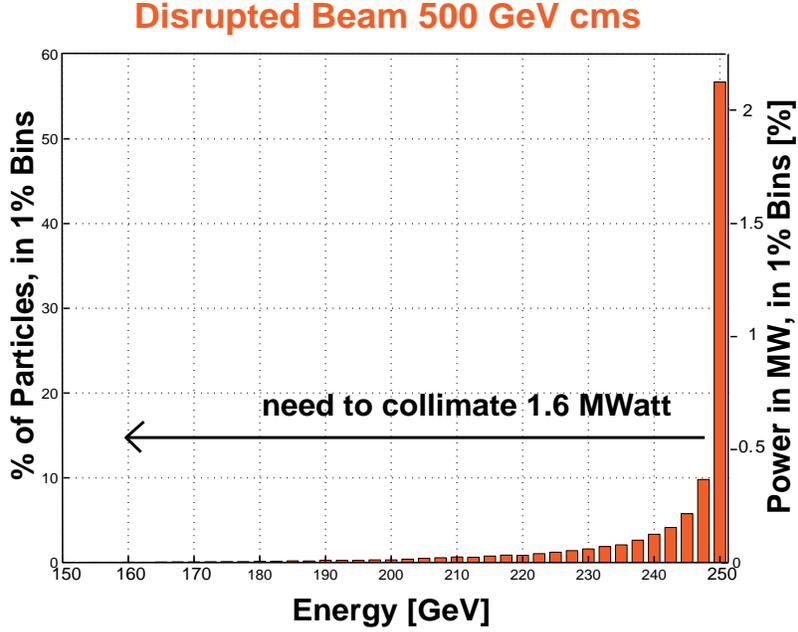,height=3.4in,width=4.2in}}
\vspace{0pt}
\caption{Histogram of particles (left ordinate) and their power (right ordinate) 
in 1\%-bins as a function of disrupted beam energy.
To achieve a $\pm$0.5\% energy resolution for 250 GeV disrupted beams, 1.6~MW need to be collimated.
}
\label{fig:power}
\end{figure}

Figure \ref{fig:power} shows the energy distribution
of the 250 GeV disrupted beam and the disrupted power in 1\% bins, calculated using the
beam-beam program GUINEAPIG\cite{guineapig}.  
In order to achieve a $\Delta E/E$ of 1\%, the quality required for fixed target experiments (see below), 1.6~MW have to be collimated in the transport line to the experimental area.
As an existence proof, the energy defining slits (SL-10) in the
SLAC A-line are designed for such power.
In the NLC case, not all power has to be absorbed by the energy defining slits because some will be absorbed by the 5~MW dump. 

At higher cms energy (1 TeV), the beams at collision are smaller, and the beamstrahlung and
coherent pair production effects more severe.  
The use of disrupted beams at higher energy also seems possible, but needs more study.  
 
\begin{figure}[bhtp] % fig 6
\centerline{\epsfig{file=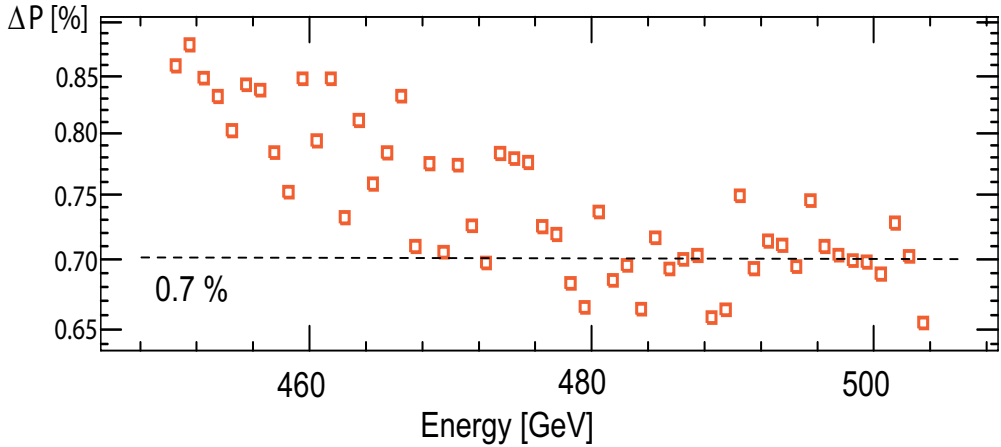,height=2.45in,width=5.2in}}
\vspace{0pt}
\caption{Depolarization as a function of disrupted beam energy. 
The data are for 1 TeV cms.
}
\label{fig:depolhist}
\end{figure}

The disrupted beam depolarization due to the beam-beam interaction at the IP was calculated using the program CAIN\cite{cain}.
Figure \ref{fig:depolhist} shows that the depolarization in the highest energy bins is small (0.7\%), even for 500 GeV beams (1 TeV cms).
The depolarization for 250 GeV beams would be 0.3\%. 

%%%%%%%%%%%%%%%%%%%%%%%%%%%%%%%%%
\section{Requirements on the Fixed Target Beams and Possible Experiments}
\subsection{Requirements on the Beam Parameters}  
 
The demands on beams for fixed target physics are very modest compared to
the demands of luminosity production with colliding beams.
For colliding beams {\bf both} beams must have a very small emittance,
and {\bf both} the electron and positron system
must be working at the highest level. 
 
These requirements do not apply for fixed target experiments.  
Here the luminosity is determined by the number of beam particles and the density of the target, and not the emittance of the beams.
Luminosities of $\approx$ 10$^{39}$ cm$^{-2}$ sec$^{-1}$ can be achieved; 
this is 4-5 orders of magnitude larger than in colliding beam experiments.
Furthermore, only the system for producing polarized electrons is needed. 

Traditionally, in preparing collider experiments a ``Snowmass year'' = 10$^7$ sec,
corresponding to 32\% efficiency, is used for run time calculations.
Because of the relaxed conditions we will use $\sqrt2$*32\% = 45\% efficiency 
for fixed target experiments at the NLC.
This number compares well with the experience at SLAC.

\vskip0pt
\begin{center}
\begin{tabular}{|l|c|}
Parameters				&Value Required			\\  \hline
Energy spread $\Delta E$	&$\leq$ 1\% E	\\  
Polarization			&$\geq$ 80\%			\\  
Beam size				&$<$$\approx$ 1 mm			\\
Charge jitter			&$<$ 2\%				\\
Position jitter on target		&$<$ 100 $\mu$m     	\\
**Beam asymmetries $\Delta Q$ 	&$\leq$ 10$^{-9}$ Q	\\
**Position asymmetries $\Delta x$ 	&$<$10 nm		\\
\end{tabular}
\end{center}
{\noindent {\footnotesize Requirements of fixed target electron beams: 
\small These requirements are similar to present day spin physics experimental needs, properties already 
achieved in End Station A at SLAC.
The two last quantities (*) are the uncertainties with respect to the polarization state integrated over an entire run.}
\vskip0pt

As shown above, the specific requirements of energy spread and 
polarization for fixed target experiments can be achieved, even with disrupted beams.
This opens up parasitic use of otherwise unused beams, thus enhancing the physics value of operations at moderate cost. 

\subsection{Fixed Target Area Design Parameters}

The fixed target area, including the targets, must be accessible during 
colliding beam operations to make installation and 
operation of the fixed target experiments efficient and independent from the colliding beam operation. 
The main colliding beam dump at $\propto$~200~m is assumed to be hermetic, except for $\mu$'s.
Muons range out with 0.4~m per GeV in earth and a transverse spread of about 10~m.
At 250 GeV they will reach $\propto$600~m in the earth beyond the main colliding beam dump,
so the transverse separation of the experimental area
from the primary line-of-flight of the $\mu$'s has to be at least 10~m. 

This optimal scattering angles at 250~GeV are 1-4 mrad for M\o ller experiments and  $\approx$ 20~mrad for spin structure experiments.
This in turn requires large drift sections to have enough transverse space available for the spectrometers. 
The target and spectrometer locations can be 
separated by a divergent tunnel to minimize the excavation costs of an experimental hall. 

\subsection{Possible Experiments}
Below is a list of possible experiments which should continue to be interesting ten years from now, thereby
opening a window of opportunity for NLC.
The availability of high energy, high flux and high polarization $\gamma$-beams will spawn new and exciting proposals because the figure of merit of $\gamma$-absorption experiments increases with energy. 

\noindent {\bf Experiments with (e,e$^\prime$):}
\begin{itemize}
\item A$_{LR}$ by M\o ller scattering (sin$^2$$\Theta$$_W$)
\item Spin structure functions at very low $x$
\end{itemize}
\noindent {\bf Experiments with $\gamma$-absorption :}
\begin{itemize}
\item Polarized gluon density  $\Delta$G
\item Gerasimov-Drell-Hearn sum rule (GDH)
\item General charm physics
\end{itemize}

\noindent With a backscattered laser beam, quasi-monochromatic $\gamma$'s of 200 GeV energy can also be produced,
but at much lower intensities.

As listed above, with $\gamma$-absorption, the gluon spin contribution $\Delta$G  in the nucleon can be determined.
This is an interesting possibility via open charm production from $\gamma$g fusion, because the cross section for $\gamma$N -$>$ c\=cX at 200 GeV is 700 nb, so that
the event rate with a primary $\gamma$-beam will be very high. 
An experiment using muons, but similar in kinematics with a much lower event rate, is COMPASS at CERN \cite{compass}.
A new proposal to do such measurements at SLAC with a 45 GeV coherent bremsstrahlung beam
(SLAC-Proposal E-161, \cite{slace161}) has just been approved.

Another experiment which could have a very large impact on the physics of charm mixing, is
$\gamma$N -$>$ c\=cX. Here the cross section is 3 orders of magnitude larger than at B-factories.  
The kinematic boost is large compared even to asymmetric B-factories, opening up a new regime for lifetime determinations, D$^0$-mixing and CP-violation. 
This experiment would improve on the work of the FOCUS collaboration (FermiLab E831, \cite{focus}) which used a 250 GeV endpoint broad band bremsstrahlung beam from secondary electrons and positrons produced from the 800 GeV Tevatron proton beam.

\subsection{Specific Example: The Running of sin$^2$$\Theta$$_W$ by Polarized M\o ller Scattering}

SLAC is currently finishing construction of a 50 GeV fixed target experiment to measure the
weak mixing angle \cite{Kumar95} at low momentum transfer (off the Z) with M\o ller scattering 
with great precision (E-158, \cite{slace158}).
The first runs are scheduled for early 2001.

Figure \ref{fig:Moller} shows a theoretical prediction and several experimental measurements of sin$^2$$\Theta$$_W$ as well as the predicted errors for two potential NLC fixed target experiments.
Like other Weinberg angle measurements \cite{SLD}, they are based on measuring A$_{LR}$, the left-right asymmetry.

\vskip10pt
\begin{figure}[bhtp] % fig 7
\centerline{\epsfig{file=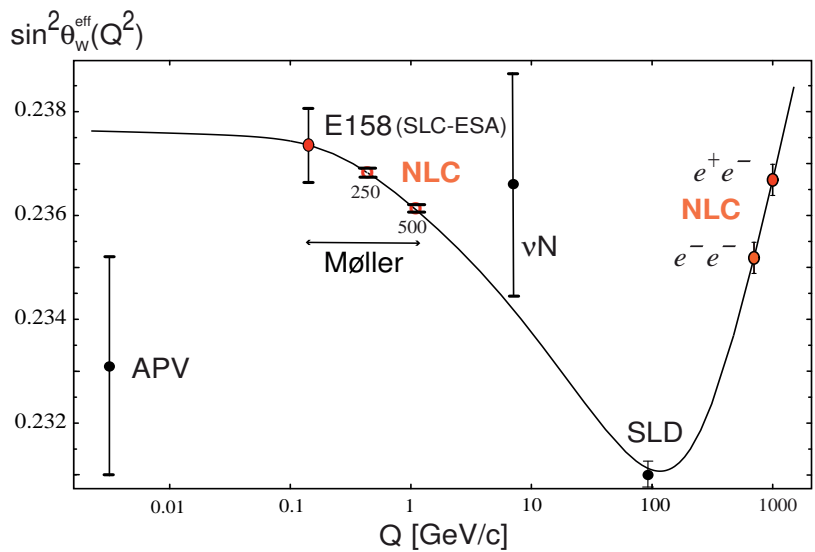,height=2.9in,width=4.2in}}
\vspace{10pt}
\caption{The prediction of sin$^2$$\Theta$$_W$ due to radiative corrections, Ref.~[18].
The data points labeled APV, $\nu$N, and SLD are experimental points, the others (red) are predictions and, therefore, are lying on the curve.
The NLC colliding beam points and errors are from Ref.~[19] and the NLC M\o ller points and errors are from Ref.~[20].
}
\label{fig:Moller}
\end{figure} 

The table below shows the effect of beam energy on the M\o ller measurement error.

\vskip0pt
\begin{center}
\begin{tabular}{|l|c|c|c}
Experiment			&E-158	&NLC-I	&NLC-II		\\  \hline
E/GeV				&46.4		&250		&500			\\  
A$_{LR}$/10$^{-7}$	&3.2		&16.1		&32.2			\\	\hline
relative $\delta$ from A$_{LR}$ alone	&1	&1/5.4	&1/10.8		
\end{tabular}
\end{center}
\noindent {\footnotesize TABLE: Impact of beam energy on the M\o ller measurement error: 
as explained in the text the statistical error decreases with increasing beam energy.
100\% polarization was assumed for the table.}

The small errors result mainly from the fact that the left-right asymmetry is proportional to the beam energy:  A$_{LR} \propto $E.
This has a unique and beneficial consequence for M\o ller scattering.
Generally in electron scattering the cross section is proportional to the inverse of the beam energy squared, $\sigma \propto E^{-2}$, but in the case of M\o ller scattering the cross section is $\propto E^{-1}$ (see Ref. \cite{slace158}).
The figure of merit for this experiment, however, is $A^2\cdot\sigma \propto E$.
The consequence is that the statistical error decreases with increasing beam energy.
The point in Figure 7 at NLC-250 assumes a total charge of 170 Coul on target.  
Assuming the nominal NLC beam of 10$^{12}$ electrons per bunch train, 
this corresponds to eight months running at 45\% efficiency using an undisrupted beam.
For the disrupted beam, 57\% of the beam is within $\Delta$E/E = 1\%.
In this case 120 Coul can be collected in nine months running, or 170 Coul in 13 months.

In summary for M\o ller scattering, the use of disrupted beams is possible
and the beam depolarization due to colliding beams is small.
Within about one year of running at NLC energies, the compositeness of the electron can be probed at the 60 TeV scale and the existence of an extra Z-boson, Z$^\prime$, can be verified or excluded up to 2.7 GeV.
An important cross check on the Higgs mass can be performed at the 10\% level \cite{kumar96}.

\section{Summary}
Single beam use at the NLC offers possibilities for a wide variety of fixed target experiments at (relatively) modest cost. 
We have investigated  the possibilities of attaching a single beam facility
to the low energy IR at 250 GeV maximum energy for polarized electrons, 200 GeV for polarized quasi-monochromatic photons.  
Other locations are possible.
NLC detector development and testing, in particular for a detector in the HE IR, is possible with the early availability of a
250 GeV test beam.

{\bf\noindent
{Acknowledgments}}\\
\noindent We would like to thank: Peter Bosted (University of Massachusetts, Amherst), 
David Burke (SLAC), Don Crabb (University of Virginia), 
{Krishna} Kumar (University of Massachusetts, Amherst), Mike Olson (St.~Norbert College), 
Tom Markiewicz (SLAC), 
Makis Petratos (Kent State University) and Terry Toole (American University), 
for much and diverse help and contributions.


\begin{references}


\bibitem{slace142} E-142 Collaboration, P.L. Anthony, ``Deep Inelastic Scattering of Polarized Electrons by Polarized $^3$He and the Study of the Neutron Spin Structure'', Phys.Rev. {\bf D54}:6620-6650,1996;~``Determination of the Neutron Spin Structure Function'',  Phys.Rev.Letters {\bf71}:959-962,1993.

\bibitem{slace143} E-143 Collaboration, K. Abe et al., ``Measurements of R=$\sigma_L$/$\sigma_T$ for 0.03$<$x$<$0.1 and Fit to World Data'', Physics Letters {\bf B452}:194-200,1999;
``Measurements of the Proton and Deuteron Spin Structure Functions g$_1$ and g$_2$'', Phys.Rev.{\bf D58}:112003-112061, 1998;
``Measurement of the Proton and Deuteron Spin Structure Function g1 in the Resonance Region'', 
Phys.Rev.Letters {\bf78}:815-819,1997;
``Measurement of the Q$^2$-Dependence of the Proton and Deuteron Spin Structure Functions 
g$^p$$_1$ and g$^d$$_1$'', Physics Letters {\bf B364}:61-68,1995;
``Measurements of the Proton and Deuteron Spin Structure Function g$_2$ and Asymmetry A$_2$'',
Phys. Rev. Letters {\bf76}:587-591,1996;
``Precision Measurement of the Deuteron Spin Structure Function g$^d$$_1$'', 
Phys.Rev.Letters {\bf75}:25-28,1995.  

\bibitem{slace154} E-154 Collaboration, K. Abe et al., 
``Measurement of the Neutron Spin Structure Function g$^n$$_2$ and Asymmetry A$^n$$_1$'', 
Physics Letters {\bf B404}:377-382,1997;
``Next to Leading Order QCD Analysis of Polarized Deep Inelastic Scattering Data'', 
Physics Letters {\bf B405}:180-190, 1997;
``Precision Determination of the Neutron Spin Structure Function g$^n$$_1$'', 
Phys.Rev.Letters {\bf 79}:26-30,1997. 

\bibitem{slace155} E-155 Collaboration, P.L. Anthony et al., 
``Measurement of the Deuteron Spin Structure Function g$^d$$_1$(x) for 
1~(GeV/c)$^2$$<$Q$^2$$<$40 (GeV/c)$^2$~'', Physics Letters {\bf B463}:339-345, 1999; 
``Inclusive Hadron Photoproduction from Longitudinally Polarized Protons and Deuterons'', Physics Letters {\bf B458}:536-544,1999; 
``Measurement of the Proton and Deuteron Spin Structure Functions g$_2$ and Asymmetry A$_2$'', Physics Letters {\bf B458}:529-535, 1999.  

\bibitem{slace144} E-144 Collaboration,  
C. Bamber et al., ``Studies of Nonlinear QED in Collisions of 46.6~GeV Electrons with Intense Laser Pulses'', Phys.Rev. {\bf D60}:092004:1-43,1999

\bibitem{cavalli93} E-146 Collaboration: M. Cavalli-Sforza et al., 
``A Method of Obtaining Parasitic $e^+$ or $e^-$ Beams during SLAC Linear Collider Operations'', SLAC-PUB-6387, 1993.

\bibitem{LPM} E-146 Collaboration: P.L. Anthony et al., 
``Bremsstrahlung Suppression due to the LPM and Dielectric Effects in a Variety of Materials'', SLAC-PUB-7413, LBL-40054, LBNL-40054, Feb 1997. 52pp. 
Published in Phys.Rev.D56:1373-1390,1997.

\bibitem{glast96} P. Anthony, ``GLAST Beam Test at SLAC'', SLAC-PUB-7323, Oct 1996. 21pp. 
Presented at Workshop on the Next Generation of High-Energy Gamma Ray Telescopes: 
Exploring the Astrophysics of Extremes, Greenbelt, MD, 4-6 Sep., 1996.

\bibitem{nlczdr} C. Adolphsen et al.,  ``Zeroth Order Design Report for the Next Linear Collider'', 
SLAC-Report {\bf 474} (1996).

\bibitem{fluka} A. Fasso, A. Ferrari, P.R. Sala and J. Ranft, ``FLUKA: Status and Prospective for Hadronic Applications'', A. Fasso, A. Ferrari, P.R. Sala,
``Electron-Photon Transport in FLUKA: Status'', Proc. of Conf. MC2000 (Advanced Monte Carlo
for Radiation Physics, Particle Transport Simulation and Applications) Lisbon, 2000.

\bibitem{guineapig} D.Schulte, Ph.D. Thesis (University of Hamburg, 1996); TESLA-97-08. 

\bibitem{cain} K.Yokoya, ``User's Manual of CAIN'', 1997.

\bibitem{slace161} http://www.slac.stanford.edu/exp/e161/

\bibitem{compass} http://wwwcompass.cern.ch/ 

\bibitem{focus} S.~Bianco, ``New FOCUS Results on Charm Mixing and CP Violation,''
hep-ex/0011055, and http://www-focus.fnal.gov.

\bibitem{Kumar95} K.S. Kumar, E.W. Hughes, R. Holmes and P.A. Souder,
``Precision Low-energy Weak Neutral Current Experiments'', Mod. Phys. Lett.{\bf A10}:2979,1995.

\bibitem{slace158} R. Carr et al., ``A Precision measurement of the Weak Mixing Angle in M\o ller Scattering'',  SLAC-Proposal-{\bf E-158}, 1997.

\bibitem{czarnecki00} A. Czarnecki and W. Marciano, ``Polarized M\o ller Scattering Asymmetries'', 
BNL-HET-00/2, hep-ph/0003049, Int.J.Mod.Phys.{\bf A15}:2365-2376,2000.

\bibitem{marciano00}  W. Marciano, ``Precision Electroweak Parameters and the Higgs Mass'', 
BNL-HET-00/04, hep-ph/0003181.

\bibitem{SLD} T. Abe, ``The Final SLD Results for A$_{LR}$ and A$_{lepton}$'', 
SLAC-PUB-8646, 2000, Invited talk presented at the 30th International Conference on High Energy Physics, 
Osaka, Japan, July 27 - August 2, 2000.

\bibitem{kumar96} Krishna Kumar, ``Fixed Target M\o ller Scattering at the NLC'', 
in Snowmass 96, new Directions for High Energy Physics, American Physical Society, 1996.

\end{references}
\end{document}